\documentclass{article}

\usepackage{arxiv}

\usepackage[utf8]{inputenc} 
\usepackage[T1]{fontenc}    
\usepackage[colorlinks=true,
            citecolor=blue,
            linkcolor=blue,
            urlcolor=cyan]{hyperref}
\usepackage{url}            
\usepackage{booktabs}       
\usepackage{bm}
\usepackage{amsfonts}       
\usepackage{amsmath}
\usepackage{amssymb}
\usepackage{mathtools}
\usepackage{authblk}
\usepackage{romannum}
\usepackage{nicefrac}       
\usepackage{microtype}
\usepackage{cleveref}
\usepackage{xcolor}
\usepackage{algpseudocode}
\usepackage{algorithm}
\usepackage{color,soul}
\usepackage{caption}
\usepackage{tabularx}
\usepackage{lipsum}
\usepackage{fancyhdr}       
\usepackage{graphicx}       
\graphicspath{{media/}}     
\usepackage{enumitem}
\makeatletter
\newcommand{\printfnsymbol}[1]{%
  \textsuperscript{\@fnsymbol{#1}}%
}

\makeatother

\definecolor{elastomer}{HTML}{FF6B6B}
\definecolor{gel}{HTML}{90E0B0}

\usepackage[normalem]{ulem}

\definecolor{DLblue}{RGB}{0,0,180}

\newcommand{\blfootnote}[1]{%
  \begingroup
  \renewcommand{\thefootnote}{}
  \footnotetext{#1}
  \endgroup
}

\title{CROPS: Cosserat-Rod-Based Co-Design of Soft Slender Programmable Material Systems for Biomedical Applications}

\author[1]{Aaditya Chandrasekhar}
\author[1]{Dex Doksoo Lee}
\author[1]{Hyunwoo Kwon}
\author[1]{Shicheng Li}
\author[1,*]{Wei Chen}

\affil[1]{Department of Mechanical Engineering, Northwestern University, Evanston, IL, USA
\authorcr
   \{\tt aadityacs, weichen\}@northwestern.edu,
   \{\tt dslee, hyunwookwon2030,shichengli2029\}@u.northwestern.edu}

\affil[*]{Corresponding author}

\begin{document}
\maketitle

 \blfootnote{The Python source code is available at \href{https://github.com/ideal-nu/crops}{github.com/ideal-nu/crops}}.

\begin{abstract}
Slender structures made of stimuli-responsive materials represent an emerging class of programmable material systems, with applications in soft robotics, minimally invasive surgery, wearable electronics, and deployable structures. In biomedical applications, the on-demand design of slender structures, encompassing both their material layout and actuation, is essential to harnessing their large deformation and multiphysics actuation to meet patient-specific design requirements. However, owing to geometric and material nonlinearities, their structure-actuation-response relationship is highly nonintuitive and computationally demanding. This often makes trial-and-error and conventional design approaches challenging for addressing patient-specific needs. To this end, we present a Cosserat-rod-based design optimization framework for the autonomous design of programmable slender structures. Leveraging the Cosserat rod representation, the end-to-end differentiable framework is demonstrated on two case studies: (i) design of swellable stents that morph to the shape of carotid vessels; (ii) co-design of the remnant magnetization and the applied field of magnetically-active catheters to reduce the force exerted on the patient-specific vessel wall during insertion. Both self-contact and external contact with surrounding structures are accounted for. We consider two actuation mechanism, namely the swelling of gel-elastomer composites and the magnetic actuation of hard-magnetic composites. The proposed framework presents an autonomous design platform of soft slender material systems that support large deformations and multiphysics actuation toward personalized biomedical design.

\end{abstract}


\section*{Highlights}
\begin{itemize}
    \item We present a differentiable design optimization framework for stimuli-responsive slender structures.
    \item Cosserat rod modeling of slender structures accommodates large deformations, multiphysics actuation, and contact.
    \item Spatially resolved material distributions and actuation variables are jointly optimized for swelling in gel-elastomer composites and magnetic actuation in hard-magnetic soft materials.
    \item Patient-specific stent and catheter designs demonstrate the capabilities of the proposed framework.
\end{itemize}

\section{Introduction}
\label{sec:intro}

\begin{figure}[h!]
    \begin{centering}
        \includegraphics[scale=0.48,trim={0 0 0 0},clip]{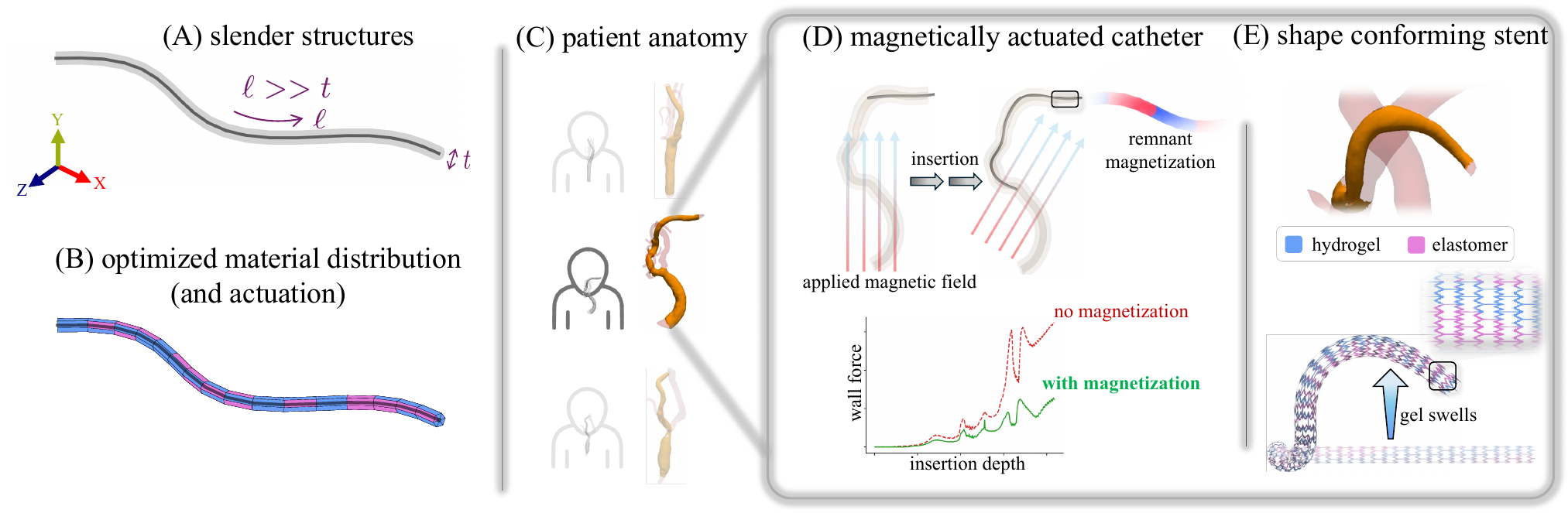}
        \caption{Overview of the proposed framework, CROPS (\underline{C}osserat \underline{R}od \underline{O}ptimization of \underline{P}rogrammable \underline{S}lender structures). Given (A) a slender structure, represented as a Cosserat rod, the framework optimizes (B) the material distribution along the length and across the cross-section of the rod, together with the applied actuation. The framework is demonstrated on two devices designed for (C) patient-specific vessel anatomy. In (D) a magnetically actuated catheter, the remnant magnetization and the applied magnetic field are co-designed, reducing the force exerted on the vessel wall during insertion in comparison to an unmagnetized catheter. In (E) a shape-conforming stent, the distribution of the hydrogel and elastomer phases is optimized such that a stent that is straight when dry morphs to the shape of the vessel upon swelling.}
        \label{fig:fig_method_graphicalAbstract}
    \end{centering}
\end{figure}

Programmable material systems (PMS) form a class of material systems that can adapt their shapes or properties in response to external stimuli, enabling dynamic changes in shape, motion, properties, or functions mediated by the stimuli. 
PMS have been developed for a range of functions, including mechanical actuation and locomotion~\cite{liwei2025MagneticMPM,dong2022untethered,hu2018locomotion,kim2019ferromagneticSoftRobotCatheter}, tuning of mechanical properties~\cite{sim2025metamaterial,zhao2023encoding}, and shape programming and reconfiguration~\cite{boley2019mutlimateriallattice,sydney2016biomimetic,kim2018ferrosoftmaterial}. Across these applications, functional behavior often emerges from nonlinear coupling among spatial material distribution, geometry, and actuation. Optimizing these variables independently or fixing some a priori may overlook design combinations that achieve better mechanical performance, motivating their co-design~\cite{leecodesign2026}. Co-design can therefore expand the range of achievable responses while better satisfying geometric and functional requirements; however, exploring this coupled design space remains computationally demanding because it requires repeated evaluations of nonlinear mechanical responses throughout design optimization.

For PMS composed of active slender structures, Cosserat rod theory~\cite{gazzola2018CosseratRods} offers a promising approach to reducing the computational burden of co-design.
Slender structures serve as versatile building blocks for PMS, enabling complex three-dimensional shape changes in applications ranging from shape-morphing lattice structures~\cite{weeger2016smerod,boley2019mutlimateriallattice} to steerable soft continuum robots~\cite{kim2019ferromagneticSoftRobotCatheter,wang2021evolutionaryDesignCatheter}. In conducting the analysis and design of such structures, Cosserat rod theory represents slender structures using centerline positions and cross-sectional orientations along their length~\cite{gazzola2018CosseratRods}. Such formulations have been applied to modeling large deformations~\cite{azad2023optimal,gazzola2018CosseratRods}, swelling-driven shape changes~\cite{cobo2025finite}, and interactions involving contact and friction~\cite{xun2024cosserat,goicoechea2019non}. 
Cosserat rod-based models can reduce the computational cost of repeated design evaluations involving nonlinear finite element analyses, complex boundary conditions, and demanding functional requirements across states, making iterative material--geometry--actuation co-design more tractable than the standard three-dimensional finite element methods developed for general purposes.

A computationally efficient co-design framework holds significant promise for biomedical devices such as catheters and stents. Their designs must accommodate patient-specific anatomical constraints while satisfying coupled mechanical, fluidic, and other functional requirements, subject to stringent safety and regulatory standards. As such, existing biomedical design approaches have explored different subsets of this coupled design space, with many focusing on programming mechanical performances. For stents, gradient-based optimization has been used to tailor mechanical stiffness and flexibility through strut dimensions, junction positions, and periodic unit-cell topology~\cite{vcanic2022optimal,vcanic2023stentCrossSection,carbonaro2023design}. For magnetically actuated catheters, evolutionary optimization has been utilized to tailor axial magnetic-particle distributions under prescribed geometry and actuation conditions~\cite{wang2021evolutionaryDesignCatheter}. More integrated approaches have explored sequential optimization of catheter geometry and material distribution~\cite{liu2025unified}, as well as joint optimization of magnetization profiles and applied magnetic fields~\cite{pittiglio2022patientSpecificCatheterMagnetic,davy2023computational}. Despite these advances, existing catheter co-design approaches remain limited to coarse axial segments or discrete material choices, rather than continuous three-dimensional vector profiles and cross-sectional layouts under contact. 
Gradient-free optimization often suffers from slow convergence and weak scalability as the number of design variables in the joint design space increases.

Addressing these limitations requires both spatially resolved material parameterizations and optimization methods that remain tractable as the number of design variables grows. Topology optimization has enabled the design of spatially heterogeneous material distributions in hard-magnetic soft materials~\cite{shelly2022Magnetic}, reprogrammable magneto-mechanical structures~\cite{zhao2023encoding}, and multi-material gel--elastomer systems~\cite{chandrasekharGELATO2026}. Differentiable simulation further enables sensitivity-based optimization of coupled material and actuation variables, as demonstrated in magnetic soft robots undergoing large deformation and contact using the material point method~\cite{liwei2025MagneticMPM}. For slender structures, gradient based approaches have been utilized to optimize centerline shapes and parameters within prescribed cross-sectional architectures using rod models with linear elastic sectional laws~\cite{weeger2022isogeometric,choi2019isogeometric}. However, extending these approaches to locally parameterized material distributions along the rod and across its cross-section requires design sensitivities that account for nonlinear coupling among material response, actuation, and deformation. Cross sectional homogenization of heterogeneous material responses incorporates local material behavior into the reduced rod model without independently resolving the three-dimensional displacement field. By differentiating through this homogenization and the nonlinear Cosserat rod solution, including contact, we demonstrate that our framework can support spatially resolved material--actuation co-design while accounting for nonlinear deformation and contact.

Building upon our previous work on differentiable multi-material topology optimization~\cite{chandrasekharGELATO2026}, we develop CROPS, a differentiable \underline{C}osserat \underline{r}od-based \underline{o}ptimization framework for \underline{p}rogrammable \underline{s}lender material systems.

The main contributions are threefold:

\begin{enumerate}
    \item \textbf{Differentiable Multiphysics Rod Formulation}:
    We develop a differentiable Cosserat rod formulation that incorporates fiber-resolved cross-sectional integration to model heterogeneous material responses under swelling and magnetic actuation. The framework further incorporates regularized contact mechanics and implicit differentiation to compute sensitivities of deformation and contact forces with respect to material and actuation variables.
    
    \item \textbf{Spatial Material--Actuation Co-Design}: We establish a gradient-based optimization framework that accommodates continuous-valued material distributions along the rod axis and across its cross-section, enabling joint optimization of spatial material properties, magnetization profiles, and applied actuation.
    
    \item \textbf{Demonstrations for Patient-Specific Biomedical Design}: We demonstrate the framework through the inverse design of gel-elastomer stents to achieve prescribed deployed geometries and co-design of magnetically actuated catheters to minimize wall-contact forces given patient-specific insertion trajectories.
\end{enumerate}


\section{Method}
\label{sec:method}

\subsection{Overview}
\label{sec:method_overview}

\begin{figure}[h!]
    \begin{centering}
        \includegraphics[scale=0.55,trim={0 0 0 0},clip]{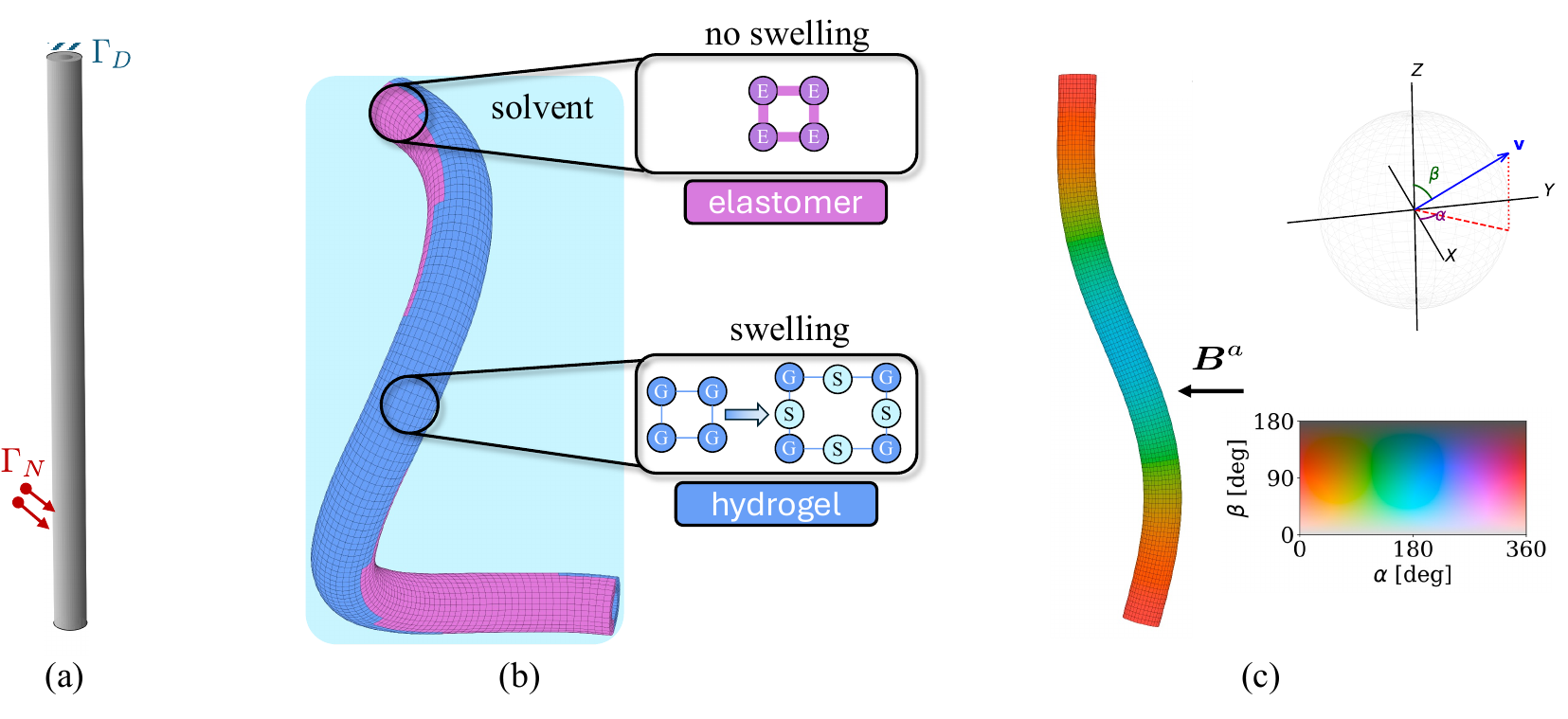}
        \caption{Actuation modes considered in this work. (a) Reference configuration of the rod with the prescribed Dirichlet $\Gamma_D$ and Neumann $\Gamma_N$ boundaries. (b) Swelling actuation of a gel-elastomer composite. Upon immersion, the hydrogel phase absorbs solvent and expands while the elastomer phase remains passive, and the resulting differential expansion drives the deformation. (c) Magnetic actuation of a hard-magnetic composite. Under the applied field $\bm{B}^a$, the embedded remnant magnetization develops moments that deform the rod. The remnant direction is parameterized by its azimuth $\alpha$ and polar angle $\beta$ in the global frame.}
        \label{fig:fig_method_overview_actuations}
    \end{centering}
\end{figure}

In this study, we consider the design optimization of programmable slender structures, in which the material distribution within the structure is optimized to obtain prescribed responses under actuation. We adopt a geometrically exact Cosserat rod \cite{simo1986CosseratRods,gazzola2018CosseratRods} to represent the slender structure. In addition to resolving the deformation through the centerline, we retain a material description over the cross-section. In particular, we consider two forms of actuation (\Cref{fig:fig_method_overview_actuations}). The first form of actuation is a gel-elastomer composite in which an active gel swells upon exposure to a solvent while a passive elastomer resists and redirects the deformation. The other one is a hard-magnetic composite in which particles carrying a remnant magnetization embedded in a soft matrix develop moments under an applied magnetic field. In both cases, the spatial distribution of material encodes the mapping from stimulus to deformation.

We begin by assuming an unstressed reference configuration of the centerline with prescribed arc length and specified cross-sectional orientations along the curve. Furthermore, we assume a prescribed set of candidate material phases and their properties, for instance the gel and elastomer phases of the swelling system. Additionally, we assume prescribed fixtures, forces, and moments. We also assume prescribed environmental conditions: (i) for swelling, these include the chemical potential, temperature, and molar volume of the solvent bath; (ii) for magnetic actuation, these include the bounds on the magnitudes of the remnant magnetization and the applied magnetic field. Where relevant, the geometry of confining surroundings and obstacles is also prescribed. Our design goal is to compute the material distribution, and where relevant the applied magnetic field, for which the actuated structure attains a prescribed target shape or mechanical output, while imposing the design constraints.

\begin{figure}[h!]
    \begin{centering}
        \includegraphics[scale=0.6,trim={0 0 0 0},clip]{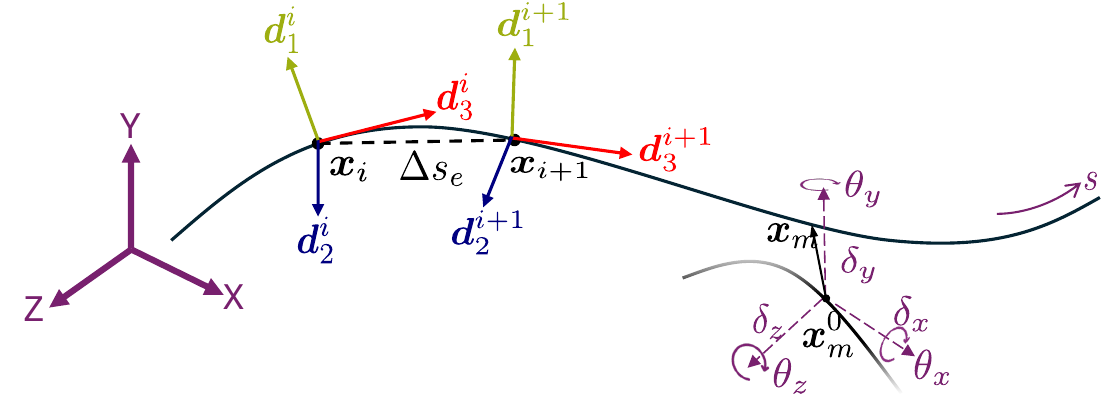}
        \caption{Discretized representation of the rod. The centerline is divided into elements of reference length $\Delta s_e$ spanning nodes $\bm{x}_i$ and $\bm{x}_{i+1}$, each node carrying a director triad $\{ \bm{d}_1^i, \bm{d}_2^i, \bm{d}_3^i \}$. The DOF are illustrated at an arbitrary node $m$, where the displacements $\delta_x, \delta_y, \delta_z$ are measured from the reference position $\bm{x}_m^0$ and the rotations $\theta_x, \theta_y, \theta_z$ carry the reference directors onto the current ones. The reference configuration is shown in gray and all global quantities in purple.}
        \label{fig:fig_method_intro_rod_centerline}
    \end{centering}
\end{figure}

We now consider the rod's centerline position $\bm{x}(s) \in \mathbb{R}^3$ parameterized by the reference arc length $s \in [0, L]$, together with an orthonormal director triad $\{ \bm{d}_1(s), \bm{d}_2(s), \bm{d}_3(s) \}$. Here, $\bm{d}_1$ and $\bm{d}_2$ span the cross-sectional plane while $\bm{d}_3$ is normal to it. The triad encodes the cross-sectional orientation, admitting bending, twist, shear, and stretch. As illustrated in \Cref{fig:fig_method_intro_rod_centerline}, the centerline is discretized into $N_e$ elements over $N_n$ nodes, with nodal positions $\bm{x}_i$, nodal triads $\{ \bm{d}_1^i, \bm{d}_2^i, \bm{d}_3^i \}$, and element $e$ spanning nodes $i$ and $i+1$ with reference length $\Delta s_e$. Each node carries three translational and three rotational degrees of freedom (DOF), $\bm{q}^{(i)} = [ \delta_x^i, \delta_y^i, \delta_z^i, \theta_x^i, \theta_y^i, \theta_z^i ]^\top$, assembled into the global state vector $\bm{q} \in \mathbb{R}^{6 N_n}$. Here, $\bm{\delta}^i$ are displacements from the reference nodal positions $\bm{x}_i^0$ and $\bm{\theta}^i$ are rotation vectors that carry the reference directors onto the current ones, both resolved in the global frame. For convenience, we collect the directors into the rotation matrix $\bm{D} \in SO(3)$, whose columns are $\bm{d}_1$, $\bm{d}_2$, and $\bm{d}_3$, and which maps material frame components onto the global frame, so that $\bm{D} = \exp ( \hat{\bm{\theta}} ) \bm{D}_0$, with $\hat{\bm{\theta}}$ the skew-symmetric matrix of $\bm{\theta}$ and $\bm{D}_0$ the reference director frame.

\begin{figure}[h!]
    \begin{centering}
        \includegraphics[scale=0.6,trim={0 0 0 0},clip]{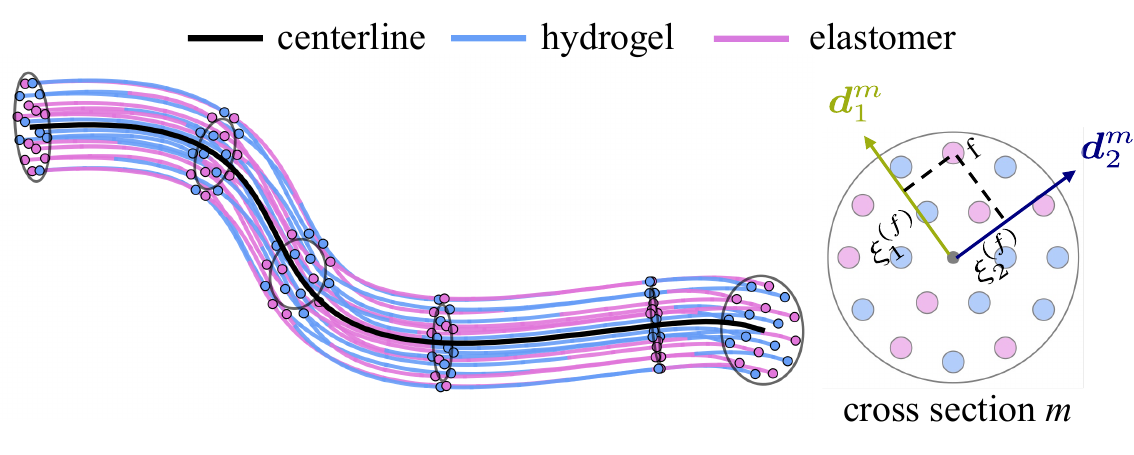}
        \caption{Fiber-resolved representation of the cross-section, illustrated for a gel-elastomer composite. Left, the fiber tracks swept along the length of the rod and colored by material phase, with cross-sections shown at selected stations. Right, a single cross-section of node $m$, where a fiber $f$ is located at the fixed material offset $\bm{\xi}^{(f)} = ( \xi_1^{(f)}, \xi_2^{(f)} )$ in the $(\bm{d}_1^m, \bm{d}_2^m)$ basis.}
        \label{fig:fig_method_fiber}
    \end{centering}
\end{figure}

Furthermore, we let the material vary both along the length and within the cross-section. Towards this, we resolve the cross-section into discrete fibers at prescribed offsets from the centerline, as illustrated in \Cref{fig:fig_method_fiber}. A fiber $f$ is located at a fixed material offset $\bm{\xi}^{(f)} = ( \xi_1^{(f)}, \xi_2^{(f)} )$ in the $(\bm{d}_1, \bm{d}_2)$ basis. We refer to the portion of fiber $f$ within element $e$ as a fiber segment, where each fiber segment carries its own material. It is the material assigned to the fiber segments that we seek to design.

We consider the finite deformation elastic response of the structure together with the swelling or magnetic actuation, the contact with the surroundings and with itself, and any applied mechanical loads. The equilibrium configuration is obtained by rendering the total potential energy stationary with respect to the degrees of freedom. We adopt gradient-based optimization, with the sensitivities computed via automatic differentiation (AD) through our end-to-end differentiable pipeline.

The remainder of this section is organized as follows. \Cref{sec:method_kinematics} presents the kinematics and the elastic response. \Cref{sec:method_swelling} and \Cref{sec:method_magnetic} describe the swelling and magnetic actuation, respectively, and \Cref{sec:method_contacts} the contact and applied loading. The design parametrization and the optimization problem are presented in \Cref{sec:method_optimization}, followed by the sensitivity analysis in \Cref{sec:method_sensAnalysis} and the complete algorithm in \Cref{sec:method_algorithm}.

\subsection{Kinematics and Elastic Response}
\label{sec:method_kinematics}

Having established the state representation, we now define the strain measures that characterize the deformation of the rod. We present a condensed account here, and refer the reader to \cite{simo1986CosseratRods, gazzola2018CosseratRods} for details.

We begin by defining the curvature $\bm{\kappa}(s)$, which describes the rate of change of the director triad along the arc length $s$ through $\partial_s \bm{d}_a = \bm{\kappa} \times \bm{d}_a$ for $a = 1, 2, 3$. Resolved in the material frame, $\kappa_1$ and $\kappa_2$ measure bending about the two cross-sectional axes, while $\kappa_3$ measures twist. Similarly, the centerline tangent resolved in the material frame defines the stretch and shear vector $\bm{\nu}(s)$, with components $\nu_a = \partial_s \bm{x} \cdot \bm{d}_a$. Here, $\nu_1$ and $\nu_2$ measure the transverse shear of the cross-section, while $\nu_3$ measures the axial stretch. Together, $(\bm{\nu}(s), \bm{\kappa}(s))$ characterize the centerline deformation. Finally, we note that the reference strains $(\bm{\nu}^0, \bm{\kappa}^0)$ follow from the prescribed reference centerline and directors, and are in general nonzero.

Having described the kinematics of the centerline, we now turn to the individual fibers. Recall that a fiber $f = 1,\ldots, N_f$ is located at a \textit{fixed} material offset $\bm{\xi}^{(f)} = (\xi_1^{(f)}, \xi_2^{(f)})$ in the $(\bm{d}_1, \bm{d}_2)$ basis (\Cref{fig:fig_method_fiber}). Fiber attributes vary both along the arc length and across the cross-section, and are evaluated at the fiber segment of element $e$ and fiber $f$ as $(\cdot)_{ef}$. We suppress these indices wherever no ambiguity arises. Furthermore, assuming the cross-section deforms rigidly in its own plane, a fiber occupies the position $\bm{x}(s, \bm{\xi}^{(f)}) = \bm{x}(s) + \xi_1^{(f)} \bm{d}_1(s) + \xi_2^{(f)} \bm{d}_2(s)$. Observe that the fiber positions are determined entirely by the centerline and the director triad. In other words, the fiber bundle introduces no DOF beyond those in $\bm{q}$. Differentiating along the arc length and resolving in the material frame yields the total fiber stretch $\tilde{\lambda} = \nu_3 - \xi_1 \kappa_2 + \xi_2 \kappa_1$ and the two fiber shears $\gamma_1 = \nu_1 - \xi_2 \kappa_3$ and $\gamma_2 = \nu_2 + \xi_1 \kappa_3$. We denote the reference stretch and shear components by $\tilde{\lambda}_0$, $\gamma_{1,0}$, and $\gamma_{2,0}$, respectively. Then, the local deformation gradient at the fiber, resolved in the material frame, can be written as:

\begin{equation}
  \bm{F} =
  \begin{bmatrix}
    \lambda_t & 0 & \eta_1 \\
    0 & \lambda_t & \eta_2 \\
    0 & 0 & \lambda
  \end{bmatrix} ,
  \label{eq:deformation-gradient}
\end{equation}

where $\lambda = \tilde{\lambda} / \tilde{\lambda}_0$, $\eta_1 = ( \gamma_1 - \gamma_{1,0} ) / \tilde{\lambda}_0$, and $\eta_2 = ( \gamma_2 - \gamma_{2,0} ) / \tilde{\lambda}_0$. Furthermore, $\lambda_t$ is the transverse stretch, taken as isotropic within the cross-sectional plane. Following from the definition of volume ratio $J = \det \bm{F} = \lambda_t^2 \lambda$, we have $\lambda_t = \sqrt{\frac{J}{\lambda}}$.

Having established the fiber kinematics, we now specify the elastic response. We adopt a neo-Hookean strain energy density
\begin{equation}
  \psi_{\mathrm{ela}} = \frac{G}{2} \left( I_1 - 3 - 2 \ln J \right) ,
  \label{eq:neohookean}
\end{equation}
where $G$ is the shear modulus and $I_1 = \mathrm{tr} ( \bm{F}^{\mathsf{T}} \bm{F} )$ is the first invariant of the right Cauchy-Green tensor. Furthermore, we assume molecular incompressibility, such that the volume ratio $J$ is prescribed by the swelling state of the fiber, with $J = 1$ in the absence of swelling (\Cref{sec:method_swelling}). The elastic response is one of several contributions to the free energy density of a fiber, which in general we write as
\begin{equation}
  \psi = \psi_{\mathrm{ela}} + \psi_{\mathrm{gel}} + \psi_{\mathrm{mag}} ,
  \label{eq:fiber-energy}
\end{equation}
where the swelling and magnetic contributions, $\psi_{\mathrm{gel}}$ and $\psi_{\mathrm{mag}}$, are introduced in \Cref{sec:method_swelling} and \Cref{sec:method_magnetic} respectively, and vanish where the corresponding actuation is absent.

The internal energy follows by integrating this density over the cross-section and along the arc length. The cross-sectional integral is evaluated as a weighted sum over the fibers, with the reference fiber area $A_{ef}$ serving as the quadrature weight, so that
\begin{equation}
  \Pi_{\mathrm{int}}(\bm{q}) = \sum\limits_{e=1}^{N_e} \Delta s_e \sum_{f=1}^{N_f} A_{ef} \, \psi_{ef} .
  \label{eq:internal-energy}
\end{equation}

The total potential energy comprises this internal energy together with the contact energy $\Pi_{\mathrm{con}}$ and the work of the applied loads $\Pi_{\mathrm{ext}}$ of \Cref{sec:method_contacts}, so that $\Pi(\bm{q}) = \Pi_{\mathrm{int}}(\bm{q}) + \Pi_{\mathrm{con}}(\bm{q}) + \Pi_{\mathrm{ext}}(\bm{q})$. The structure is in equilibrium when this energy is stationary with respect to the DOF. We therefore define the residual
\begin{equation}
  \bm{\mathcal{R}}(\bm{q}) = \frac{\partial \Pi}{\partial \bm{q}} ,
  \label{eq:residual}
\end{equation}
and the equilibrium configuration $\bm{q}^{\star}$ is the solution of $\bm{\mathcal{R}}(\bm{q}^{\star}) = \bm{0}$ subject to the prescribed boundary conditions. This system is nonlinear in $\bm{q}$ and is solved iteratively. At each iteration the tangent stiffness $\bm{K}(\bm{q}) = \partial^2 \Pi / \partial \bm{q}^2$ is assembled and the increment $\Delta \bm{q} = - \bm{K}^{-1} \bm{\mathcal{R}}$ is applied, until the residual falls below a prescribed tolerance. Given the nonlinearity of the response, the actuation is applied incrementally over $K$ load steps, with the solution at each load step initialized from that of the preceding one. We denote the equilibrium configuration at load step $k$ as $\bm{q}^{\star}_k$.

\subsection{Actuation by Swelling}
\label{sec:method_swelling}

Having established the rod kinematics, we now describe the first of the two forms of actuation, in which programmability is achieved via the distribution of an active swellable gel and a passive elastomer matrix. Here, we briefly outline the formulation, with further details provided in our earlier work \cite{chandrasekharGELATO2026}. When a gel is immersed in a solvent bath, it absorbs solvent molecules and expands, while an elastomer remains largely passive (\Cref{fig:fig_method_overview_actuations}). To quantify the swelling, we introduce a polymer volume fraction $\phi(s, \bm{\xi}) \in (0, 1]$, where $\phi = 1$ is the fully dry state and $\phi \to 0$ the fully swollen limit. Recall that molecular incompressibility prescribes the volume ratio in \Cref{eq:neohookean} as $J = 1/\phi$.

Following Flory-Rehner theory \cite{flory1943FloryRhener, hong2008TheoryCoupled}, the swelling contribution to the free energy density $\psi_{\mathrm{gel}}$ comprises the energy of mixing the polymer with the solvent and the chemical work of exchanging solvent with the bath, and is given by
\begin{equation}
  \psi_{\mathrm{gel}} = \frac{R T}{\tilde{\Omega}} \left[ \frac{1 - \phi}{\phi} \ln ( 1 - \phi ) + \chi (1-\phi) \right] - ( \mu - \mu^0 ) \, \frac{1}{\tilde{\Omega}} \left( \frac{1}{\phi} - 1 \right) ,
  \label{eq:gel-energy}
\end{equation}
where the solvent is characterized by its molar volume $\tilde{\Omega}$ and reference chemical potential $\mu^0$, the bath by its temperature $T$ and chemical potential $\mu$, and $R$ is the universal gas constant.

The material response is governed jointly by its Flory-Huggins interaction parameter $\chi$, which quantifies the affinity of the polymer network for the solvent, and the shear modulus $G$ associated with the elastic energy in \Cref{eq:fiber-energy}. A compliant gel network with low $\chi$ readily takes up solvent and swells, whereas a stiff elastomer network with high $\chi$ resists uptake and remains effectively passive. It is the contrast in $G$ and $\chi$ between the two phases that produces shape change. Gel fibers take up solvent and expand while the surrounding elastomer resists. As the resulting differential expansion drives the deformation, the distribution of the two phases determines the deformed shape upon actuation.

Finally, the polymer volume fraction is determined by the thermodynamic equilibrium condition between the gel and the external solvent, namely the stationarity of the free energy density with respect to $\phi$. Applying $\partial \psi / \partial \phi = 0$ to \Cref{eq:fiber-energy} yields the implicit Flory-Rehner relation
\begin{equation}
 \mathcal{F}(\phi) \coloneqq  \frac{\mu^0 - \mu}{R T} + \ln ( 1 - \phi ) + \phi + \chi \phi^2 + \frac{\tilde{\Omega} G}{R T} \, \phi \left( \lambda_t^2 - 1 \right) = 0 , \qquad \lambda_t^2 = \frac{1}{\phi \lambda} .
  \label{eq:flory-rehner}
\end{equation}
Observe that $\phi$ depends on the fiber stretch $\lambda$, so the swelling response is coupled to the mechanical state and is resolved together with the equilibrium configuration.

\subsection{Magnetic Actuation}
\label{sec:method_magnetic}

We now describe the second form of actuation, in which programmability is achieved via determining the distribution of oriented hard-magnetic particles within a soft matrix, together with the externally applied magnetic field. The following account builds on earlier treatments detailed in \cite{zhao2019mechanics, shelly2022Magnetic, liwei2025MagneticMPM}. We consider ideal hard-magnetic soft materials, wherein the embedded particles are of high coercivity, with the permanent magnetization imprinted during fabrication. We further assume that the permeability of the composite is that of the ambient medium, and that magnetic interactions between distinct parts of the structure are negligible. We describe the magnetized state through the remnant magnetic flux density $\bm{B}^r(s, \bm{\xi})$, specified in the global frame in the reference configuration and unaltered by the subsequent deformation of the structure. Actuation is in turn supplied by an applied magnetic flux density $\bm{B}^a$, specified in the global frame as a sequence of spatially uniform fields, one per load step (\Cref{fig:fig_method_overview_actuations}).

Following \cite{shelly2022Magnetic}, the magnetic contribution $\psi_{\mathrm{mag}}$ to \Cref{eq:fiber-energy} is given by
\begin{equation}
  \psi_{\mathrm{mag}} = - \frac{1}{\mu_0} \, \bm{b}^a \cdot \left( \bm{F} \, \bm{b}^r \right) , \qquad \bm{b}^a = \bm{D}^{\mathsf{T}} \bm{B}^a , \qquad \bm{b}^r = \bm{D}_0^{\mathsf{T}} \bm{B}^r ,
  \label{eq:magnetic-energy}
\end{equation}
where $\mu_0$ is the vacuum permeability, and $\bm{b}^a$ and $\bm{b}^r$ are the applied and remnant flux densities resolved in the current and reference material frames respectively.

Since the applied field is spatially uniform, it exerts no net force on the structure. Instead, the actuation arises from the torques that the field exerts on the magnetized fibers, which tend to align the remnant magnetization with the applied field. Varying the remnant direction across the cross-section and along the length produces spatially varying torques under the applied field. The distribution of the remnant magnetization, together with the applied field, determines the actuated response.

\subsection{Contacts and Applied Loadings}
\label{sec:method_contacts}

We now describe the interaction of the rod with its surroundings and with itself, together with the applied mechanical loading. Structures deployed within confined geometries or in the presence of obstacles can make contact with the surrounding boundary. Furthermore, structures undergoing large deformations may fold and make contact onto themselves. We consider both, referred to as external contact and self-contact, respectively. For external contact, we represent the surrounding geometry implicitly through a signed distance field $\varphi(\bm{x})$, adopting the convention that $\varphi > 0$ in free space. Furthermore, we assign to each element $e$ a contact radius $r_e$ of its cross-section.

The penetration depth of an element $e$ in external contact is $\epsilon_e = r_e - \varphi(\bm{x}_e)$, where $\bm{x}_e$ is the element center. For self-contact, the penetration depth of a pair of elements $a$ and $b$ is $\epsilon_{ab} = ( r_a + r_b ) - h_{ab}$, where $h_{ab}$ is the minimum distance between the two element segments. In either case, the penetration is inactive when $\epsilon \le 0$, and a penalty energy is accumulated otherwise. We adopt a $C^2$ continuous smooth ramp to regularize $\max ( \epsilon, 0 )$:
\begin{equation}
  \varrho(\epsilon) =
  \begin{cases}
    0 & \epsilon \le 0 \\[4pt]
    \dfrac{\epsilon^3}{w^2} - \dfrac{\epsilon^4}{2 w^3} & 0 < \epsilon < w \\[6pt]
    \epsilon - \dfrac{w}{2} & \epsilon \ge w
  \end{cases}
  \label{eq:regularized-penetration}
\end{equation}
where $w$ is a transition width taken as a fraction of the contact radius, and $\varrho$ recovers $\max ( \epsilon, 0 )$ as $w \to 0$. The $C^2$ continuity ensures that the tangent stiffness remains continuous at the onset of contact and the sensitivities of \Cref{sec:method_sensAnalysis} are well defined.

The contact energy of an element, or of a pair of elements, is quadratic in the regularized penetration, and is given by $\pi_{\mathrm{con}} = \tfrac{1}{2} \omega_{\mathrm{con}} \varrho^2 \ell_{\mathrm{con}}$, where $\omega_{\mathrm{con}}$ is a penalty stiffness and $\ell_{\mathrm{con}}$ is a representative contact length, taken as the element length for external contact and as the geometric mean of the two element lengths for self-contact. Summation over the elements and over the interacting element pairs gives $\Pi_{\mathrm{con}}(\bm{q})$. To preserve the sparsity of the tangent stiffness, the self-contact contribution to the tangent is retained only in block-diagonal form. Furthermore, the contact force at a node $i$ then follows from the contact energy as
\begin{equation}
  \bm{f}^{\mathrm{con}}_i = - \frac{\partial \Pi_{\mathrm{con}}}{\partial \bm{\delta}^i} ,
  \label{eq:contact-force}
\end{equation}
where $\bm{\delta}^i$ are the translational DOF of node $i$. In other words, $\bm{f}^{\mathrm{con}}_i$ is the force exerted on the rod at node $i$ by the surroundings or by the rod itself, and is equal and opposite to the force that the rod exerts in return.

Finally, the applied mechanical loading contributes $\Pi_{\mathrm{ext}}(\bm{q}) = - \bm{f}^{\mathsf{T}} \bm{q}$, where $\bm{f}$ collects the prescribed nodal forces and moments. We treat the magnitude and direction of the applied loads as fixed as the structure deforms.

\subsection{Optimization Formulation}
\label{sec:method_optimization}

We now integrate the preceding kinematic and actuation formulations to outline the proposed optimization framework. The numerical experiments in \Cref{sec:expts} consider different design variables, objectives, and constraints; we present the complete set here and invoke the relevant subset in each study.

\subsubsection{Design Variables}
\label{sec:method_opt_desVar}

Recall that the design comprises the material assigned to every fiber segment, together with, for magnetic actuation, the applied magnetic field at every load step. We collect these into the vector of design variables $\bm{\zeta}$. Since the underlying physical quantities differ in units and in magnitude, each design variable is normalized to $[0, 1]$. In particular, a physical quantity $p$ with a prescribed extent $[p_{\min}, p_{\max}]$ is recovered from its design variable $\zeta$ as $p = p_{\min} + \zeta ( p_{\max} - p_{\min} )$. This renders the design variables comparable in magnitude, and reduces the bounds on the physical quantities to the box constraint $\bm{0} \le \bm{\zeta} \le \bm{1}$.

For swelling actuation, a design variable is assigned to every fiber segment. The design variables are filtered and projected, as detailed in \Cref{sec:expts}, to obtain the pseudodensity $\rho_{ef} \in [0, 1]$ of every fiber segment, where $\rho_{ef} = 1$ corresponds to gel and $\rho_{ef} = 0$ corresponds to elastomer. The shear modulus $G_{ef}$ and the Flory-Huggins parameter $\chi_{ef}$ are then interpolated between those of the two phases (\Cref{eq:material_interp}).

For magnetic actuation, the design variables describe the remnant flux density $\bm{B}^r$ of the fiber segments, together with the applied flux density $\bm{B}^a_k$ at every load step $k$. Each flux density is parameterized by its magnitude, its azimuth $\alpha$, and its polar angle $\beta$ (\Cref{fig:fig_method_overview_actuations}). Here, the magnitude is bounded by a prescribed maximum, while the two angles span all unique orientations.

\subsubsection{Objectives}
\label{sec:method_opt_objectives}

\paragraph{Shape morphing :} One of the central goals of the case studies showcased herein (\Cref{sec:expts_NU,sec:expts_stent}) is computing the optimal material distribution such that the structure morphs to a target shape upon actuation. Given a target centerline curvature $\hat{\bm{\kappa}}_e$ and the realized centerline curvature $\bm{\kappa}_e(\bm{q}^*;\,\bm{\zeta})$ at the element centers in the swollen configuration, the objective is expressed as the minimization of the relative squared error between the two fields:
\begin{equation}
  \mathcal{J}(\bm{\zeta})
    = \frac{\sum\limits_{e=1}^{N_e} \| \bm{\kappa}_e(\bm{q}^*; \bm{\zeta}) - \hat{\bm{\kappa}}_e \|_2^2}{\sum\limits_{e=1}^{N_e} \| \hat{\bm{\kappa}}_e \|_2^2}.
  \label{eq:objective_morphing}
\end{equation}

\paragraph{Contact forces :} Another goal of the case studies showcased herein (\Cref{sec:expts_catheter}) is computing the remnant magnetization and the applied magnetic field such that the forces exerted on the surroundings during the actuation are minimized. Recall that the contact force at a node is given by \Cref{eq:contact-force}. Denoting by $\bm{f}^{\mathrm{con}}_{n,k}$ the contact force at node $n$ in the equilibrium configuration $\bm{q}^{\star}_k$ of load step $k$, the objective is expressed as the minimization of the contact force summed over all the nodes and all the load steps:
\begin{equation}
  \mathcal{J}(\bm{\zeta})
    = \sum\limits_{k=1}^{K} \sum\limits_{n=1}^{N_n} \| \bm{f}^{\mathrm{con}}_{n,k}(\bm{q}^{\star}_k; \bm{\zeta}) \|_2 .
  \label{eq:objective_contact}
\end{equation}

\subsubsection{Constraints}
\label{sec:method_opt_constraints}

\paragraph{Grayness constraint :} Recall that while the pseudodensities are relaxed to a continuous range ($0 \le \rho_{ef} \le 1$) to enable gradient-based optimization, physical realizability dictates that every fiber segment be composed of a single phase ($\rho_{ef} \in \{0, 1\}$). To drive the design toward this binary state and eliminate intermediate pseudodensities, we introduce a grayness constraint \cite{qian2017undercut}:
\begin{equation}
  g_{r} \coloneqq \frac{1}{N_eN_f} \sum\limits_{e=1}^{N_e}\sum\limits_{f=1}^{N_f} \rho_{ef} ( 1 - \rho_{ef} ) - \epsilon_g^* \le 0 ,
  \label{eq:grayness_constraint}
\end{equation}
where $\epsilon_g^* > 0$ is the permitted grayness, governed by a continuation scheme. Initially, $\epsilon_g^*$ is assigned a sufficiently large value to permit mixed material phases, enabling the optimizer to broadly explore the design space without prematurely trapping in local minima. Over the course of the optimization, $\epsilon_g^*$ is systematically annealed toward zero. This progressive tightening imposes a strict penalty on gray regions, effectively resolving the design to a discrete partition.

\paragraph{Gel volume constraint :} In some design scenarios, a constraint is imposed on the volume fraction of the gel, for instance to limit the usage of the active material. With $A_{ef}$ and $\Delta s_e$ being the reference area and the reference length of a fiber segment, the constraint is expressed as:
\begin{equation}
  g_{v} \coloneqq \frac{\sum\limits_{e,f} \rho_{ef} \, A_{ef} \, \Delta s_e}{\sum\limits_{e,f} A_{ef} \, \Delta s_e} - V_{gel}^* \le 0 ,
  \label{eq:gel_volume_constraint}
\end{equation}
where $V_{gel}^* \in (0, 1]$ is the prescribed upper bound on the volume fraction of the gel.

\subsubsection{Optimization Problem}
\label{sec:method_opt_problem}

Collecting the objectives (\Cref{eq:objective_morphing,eq:objective_contact}), the residual (\Cref{eq:residual}), and the constraints (\Cref{eq:grayness_constraint,eq:gel_volume_constraint}), the canonical optimization problem can be expressed as:

\begin{subequations}
	\label{eq:optimization_canonical}
\begin{align}
    \underset{\bm{\zeta}}{\text{minimize}} \quad & \mathcal{J}( \bm{\zeta}, \{ \bm{q}^{\star}_k \} ) \\
    \text{subject to} \quad & \bm{\mathcal{R}}( \bm{q}^{\star}_k ; \bm{\zeta} ) = \bm{0}, \; k = 1, \ldots, K \\
    & g_i ( \bm{\zeta} ) \le 0, \; i \in \{ r, v \} \\
    &  \bm{0} \le \bm{\zeta} \le \bm{1}
\end{align}
\end{subequations}

We employ the method of moving asymptotes (MMA) \cite{svanberg1987MMA} to perform the design updates. The choices of the MMA move limit and the step tolerance are specified later under numerical experiments. Default values are used for other hyperparameters.

\subsection{Sensitivity Analysis}
\label{sec:method_sensAnalysis}

We now detail the computation of the sensitivities of the objective and the constraints with respect to the design variables. Deriving and computing these is non-trivial in the present setting, owing to the geometric nonlinearity, the contact interactions, the coupling between deformation and swelling, and the sequence of equilibrium solves over which the response is evaluated. Towards this, we construct an end-to-end differentiable pipeline using the reverse-mode AD capabilities of JAX \cite{jax2018github}. This eliminates the laborious and error-prone derivation of sensitivity expressions, and accommodates varying objectives, constraints, and discretizations.

Most of the computational pipeline, from the design vector through the material properties to the objective, is an explicit composition of differentiable operations to which the chain rule applies directly. However, two quantities do not admit a closed-form expression in the design variables. The first is the equilibrium configuration $\bm{q}^{\star}$, which satisfies $\bm{\mathcal{R}}(\bm{q}^{\star}; \bm{\zeta}) = \bm{0}$ and is resolved via Newton-Raphson iteration. The second is the polymer volume fraction $\phi^{\star}$, which satisfies the Flory-Rehner relation of \Cref{eq:flory-rehner} and is resolved via bisection. Both are iterative root-finding procedures. Differentiating through the iteration history is prohibitive in memory and compute. We instead apply the implicit function theorem (IFT) \cite{blondel2022ImplicitDiff} uniformly to both.

For a converged solution $y^{\star}$ satisfying $\mathcal{H}(y^{\star}; \bm{\vartheta}) = 0$, and provided $\partial \mathcal{H} / \partial y$ is invertible at that state, the IFT gives
\begin{equation}
  \frac{d y^{\star}}{d \bm{\vartheta}} = - \left( \frac{\partial \mathcal{H}}{\partial y} \right)^{-1} \frac{\partial \mathcal{H}}{\partial \bm{\vartheta}} ,
  \label{eq:ift}
\end{equation}
replacing differentiation through the iteration history with two partial derivatives evaluated at $y^{\star}$. Setting $y = \bm{q}^{\star}$, $\mathcal{H} = \bm{\mathcal{R}}$, and $\bm{\vartheta} = \bm{\zeta}$, we obtain $d \bm{q}^{\star} / d \bm{\zeta} = - \bm{K}^{-1} \, \partial \bm{\mathcal{R}} / \partial \bm{\zeta}$, where $\bm{K} = \partial \bm{\mathcal{R}} / \partial \bm{q}$ is the tangent stiffness at the converged configuration. Setting $y = \phi^{\star}$, $\mathcal{H} = \mathcal{F}$, and $\bm{\vartheta} = (\lambda, G, \chi)$, we obtain a scalar division at each fiber segment, since $\partial \mathcal{F} / \partial \phi$ is nonzero for all physically realizable states. The two rules are applied hierarchically. As $\phi$ depends on the fiber stretch and hence on $\bm{q}$, its rule is invoked within the evaluation of $\partial \bm{\mathcal{R}} / \partial \bm{\zeta}$.

Finally, recall that the actuation is applied as a sequence of load steps. The sensitivity propagates through the corresponding sequence of equilibrium solves in reverse, at a cost of one linear solve per load step.

\subsection{Algorithm}
\label{sec:method_algorithm}

We assume a prescribed reference configuration of the rod, the candidate materials, the boundary conditions, the actuation, and, where relevant, the geometry of the surroundings. The procedure is as follows:

\begin{enumerate}
  \item \textbf{Rod initialization}: The reference centerline and directors are defined, and the centerline is discretized into $N_e$ elements (\Cref{sec:method_overview}). The cross-section is resolved into $N_f$ fibers with prescribed offsets and areas. The fixtures and the applied loads are prescribed.
  \item \textbf{Material and actuation initialization}: The material constants are specified. For swelling actuation, these include the shear moduli and the Flory-Huggins parameters of the gel and elastomer phases, together with the molar volume of the solvent, the temperature, and the chemical potentials of the dry and wet states. For magnetic actuation, these include the shear modulus and the bounds on the remnant and applied flux densities. Where relevant, the signed distance field of the surroundings and the contact parameters are specified (\Cref{sec:method_contacts}).
  \item \textbf{Solvers and optimizer}: The tolerances for the bisection solver and the Newton-Raphson iteration are specified, along with the number of load steps $K$. The optimization settings are then prescribed, including the filter and projection parameters, the constraint bounds, and the MMA move limit. Note that we adopt continuation schemes for the projection and the grayness constraint.
  \item \textbf{Design variable initialization}: The normalized design variables $\bm{\zeta}$ are initialized (\Cref{sec:method_opt_desVar}).
  \item \textbf{Property mapping}: At each iteration, the design variables are mapped to the physical quantities. For swelling actuation, the design variables are filtered and projected to obtain the pseudodensities $\rho_{ef}$, from which the shear modulus $G_{ef}$ and the Flory-Huggins parameter $\chi_{ef}$ of every fiber segment are interpolated (\Cref{eq:material_interp}). For magnetic actuation, the design variables are renormalized to obtain the remnant flux density of every fiber segment and the applied flux density at every load step.
  \item \textbf{Equilibrium analysis}: The actuation is applied incrementally over the $K$ load steps. At each load step, the equilibrium configuration $\bm{q}^{\star}_k$ is obtained by solving \Cref{eq:residual} via the Newton-Raphson method, starting from the configuration of the preceding load step. Within each Newton-Raphson iteration, the strains are projected onto the fiber segments to assemble the deformation gradient (\Cref{eq:deformation-gradient}). Where swelling is present, the Flory-Rehner relation (\Cref{eq:flory-rehner}) is solved for the polymer volume fraction. The free energy density (\Cref{eq:fiber-energy}) is then evaluated, with the swelling and magnetic contributions given by \Cref{eq:gel-energy} and \Cref{eq:magnetic-energy}. Finally, the internal energy (\Cref{eq:internal-energy}) is assembled together with the contact energy (\Cref{sec:method_contacts}) and the work of the applied loads, from which the residual and the tangent stiffness follow.
  \item \textbf{Objective and constraints}: The relevant objective (\Cref{eq:objective_morphing} or \Cref{eq:objective_contact}) is evaluated from the equilibrium configurations, and the relevant constraints (\Cref{eq:grayness_constraint,eq:gel_volume_constraint}) are evaluated from the pseudodensities.
  \item \textbf{Sensitivity analysis}: The sensitivities of the objective and the constraints with respect to the design variables are computed via reverse-mode AD through our end-to-end differentiable pipeline, as described in \Cref{sec:method_sensAnalysis}.
  \item \textbf{Design update}: The objective, the constraints, and their sensitivities are passed to MMA, which computes the updated design variables. The parameters governed by a continuation scheme are then advanced.
\end{enumerate}

Steps 5 through 9 are repeated until the convergence criterion is satisfied or the maximum number of iterations is reached.

\section{Numerical Studies}
\label{sec:expts}

In this section, we present three case studies to illustrate the proposed framework. All case studies are conducted on an AMD EPYC 7413 CPU using the JAX library \cite{jax2018github} in Python.
The three case studies are summarized in \Cref{tab:case_summary}, and are ordered by increasing complexity.

\begin{table}[]
  \centering
  \caption{Summary of the case studies.}
  \label{tab:case_summary}
  \small
  \begin{tabularx}{\textwidth}{@{} l X X X @{}}
    \toprule
    & \textbf{Case I} (\Cref{sec:expts_NU}) & \textbf{Case II} (\Cref{sec:expts_stent}) & \textbf{Case III} (\Cref{sec:expts_catheter}) \\
    \midrule
    Application & ``NU'' shape morphing & Patient-specific stents & Magnetically steered catheters \\
    Demonstrates & Interpretable shape morphing & Patient-specific shape morphing & Material-stimulus co-design \\
    Deformation & Planar ($x$-$z$) & 3D & 3D \\
    Cross-section & Rectangular & Annular lattice & Annular \\
    Aspect ratio $\mathcal{A}$ & $L/H = 200, 250, 300$ & $L/R = 35$ & $L/R_o = 175$ to $270$ \\
    $N_e \times N_f$ & $300 \times 20$ & $300 \times 180$ & $120 \times 8$ \\
    Actuation & Swelling & Swelling & Magnetic field \\
    Contact & - & - & Rigid, frictionless wall \\
    Design variables & $\rho_{ef}$ per fiber segment & $\rho_{ef}$ per fiber segment on lattice & $\bm{B}^r_e$ per element, $\bm{B}^a_k$ per increment \\
    Objective & Curvature mismatch (\cref{eq:objective_morphing}) & Curvature mismatch (\cref{eq:objective_morphing}) & Wall force (\cref{eq:objective_contact}) \\
    \bottomrule
  \end{tabularx}
\end{table}

Furthermore, the default parameters shared across the case studies are as follows, unless otherwise specified:

\begin{itemize}
  \item \textbf{Reference configuration}: The rod is straight in its stress-free undeformed reference configuration.
  
  \item \textbf{Design variables}: The design variables $\bm{\zeta}$ are normalized to $[0, 1]$ and initialized with 0.5.
  
  \item \textbf{Filter}: To ensure smoothness, the design variables are filtered along the rod axis using a linear density filter with a radius of $2L/N_e$.
  
  \item \textbf{Projection}: For the swelling experiments, the filtered variables are thresholded using a smoothed Heaviside projection \cite{sigmund2011projectionFilter} with the sharpness increased linearly from 1 to 128 over the first 60 iterations.
  
  \item \textbf{Materials}: All materials are modeled as neo-Hookean with molecular incompressibility. For the swelling experiments ( \Cref{sec:expts_NU,sec:expts_stent}), the gel and elastomer Flory-Huggins parameters are set to $\chi_{gel} = 0.2$ and $\chi_{elast} = 2.5$, and the shear modulus as $G_{gel} = 1$ MPa and $G_{elast} = 50$ MPa, respectively. For the magnetic experiment (\Cref{sec:expts_catheter}), the shear modulus is set to $G = 303$ kPa \cite{liwei2025MagneticMPM}.
  
  \item \textbf{Solvent}: For swelling, a solvent with molar volume $\tilde{\Omega} = 1.8 \times 10^{-5}$ m$^3$ mol$^{-1}$ at a temperature $T = 310$ K is assumed, with $\mu^{0} = 0$, $\mu_{dry} = -1 \times 10^{5}$ J mol$^{-1}$, and $\mu_{wet} = -100$ J mol$^{-1}$ (\Cref{eq:flory-rehner}). The chemical potential $\mu$ is increased geometrically from $\mu_{dry}$ to $\mu_{wet}$ over 70 load steps.
  
  \item \textbf{Nonlinear solver}: The equilibrium equations are solved using the Newton-Raphson method with Armijo backtracking to a relative residual tolerance of $10^{-8}$. A direct sparse solver is enlisted for the linear solves.
  
  \item \textbf{Grayness constraint}: For the swelling experiments, the grayness slack $\epsilon_g^*$ is decreased linearly from 0.3 to 0.02 over the first 60 iterations.
  
  \item \textbf{Optimizer}: The optimization is performed using MMA \cite{svanberg1987MMA}, and is terminated when the KKT residual or the change in the design variables falls below $10^{-6}$, or when the maximum number of 300 iterations is reached.
\end{itemize}

\subsection{Case I: Planar Shape Morphing of a Layered Rod}
\label{sec:expts_NU}

Recall that the central hypothesis of our work is that Cosserat rods are well suited for the representation, design, and optimization of slender structures. While conventional continuum FEA requires meshing the cross-section, the proposed formulation discretizes only the centerline, accounting for the fibers solely through integration over the cross-section (\Cref{sec:method_overview}). To illustrate this, we begin by considering a simple planar example, where a slender rod morphs to a target shape upon swelling (\Cref{tab:case_summary}). In this case study, the target shape upon deformation is chosen such that the optimized material layout can be interpreted by inspection.

\begin{figure}[]
    \begin{centering}
        \includegraphics[scale=0.46,trim={0 0 0 0},clip]{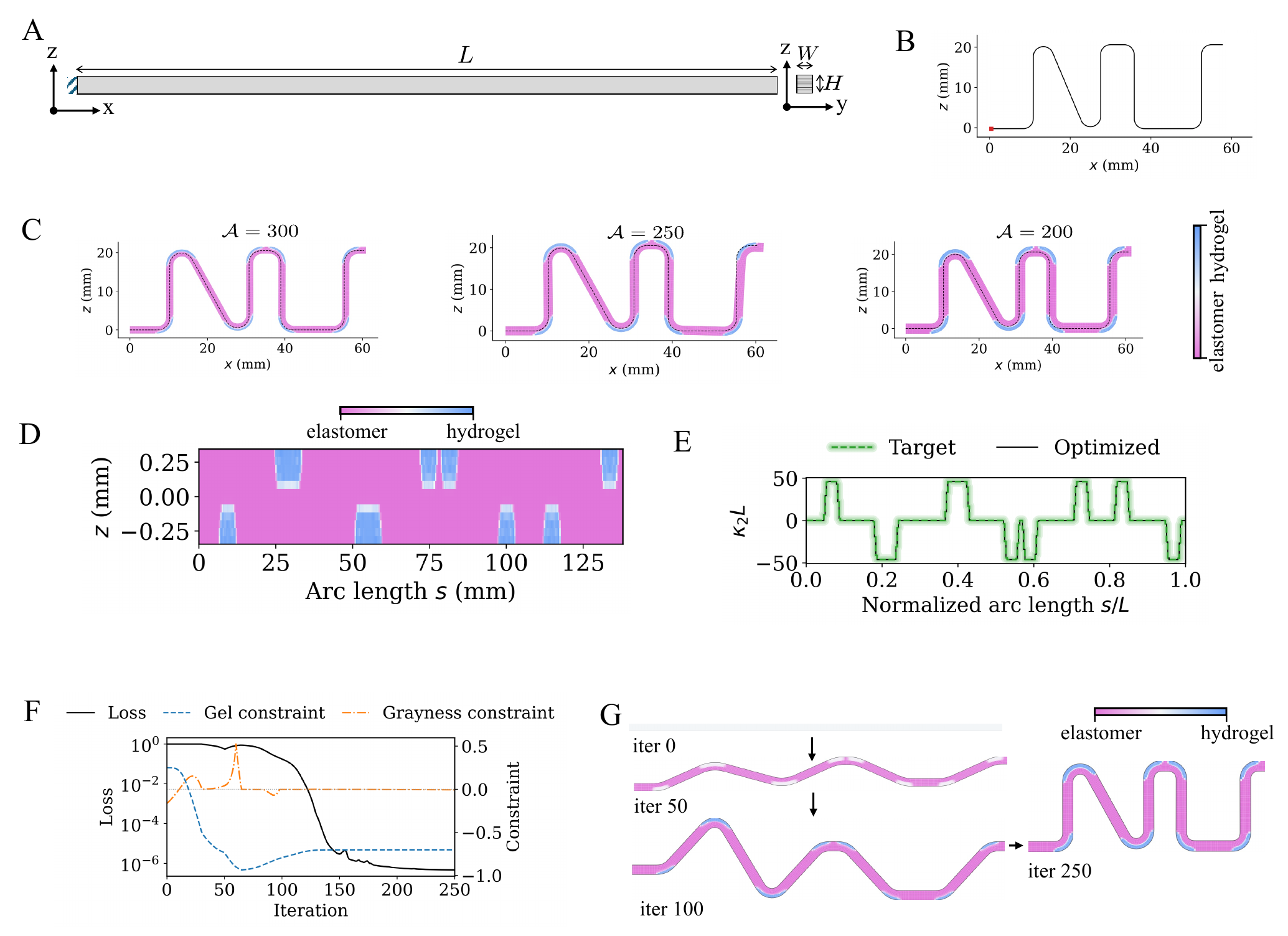}
        \caption{Planar shape morphing of a slender rod. (A) The rod and its cross-section. (B) The target shape. (C) Swollen configurations of the optimized designs for three aspect ratios, $\mathcal{A} = 300, 250, 200$. (D)-(G) show further analyses for $\mathcal{A} = 200$. (D) Optimized material layout in the undeformed configuration. (E) Realized and target curvature.  (F) Convergence of the objective and the constraints. (G) Designs at select iterations. }
        \label{fig:fig_result_swell_nu}
    \end{centering}
\end{figure}

Consider a straight rod of length $L = 138$ mm with a rectangular cross-section of width $W$ and height $H$, clamped at its left end, as illustrated in \Cref{fig:fig_result_swell_nu}(A). For illustration, we consider three slender aspect ratios, $\mathcal{A}  \coloneqq L/H = 300$, $250$, and $200$, obtained by varying $H$ with $L$ fixed and $W = 0.7H$. The rod centerline is discretized into $N_e = 300$ elements, and the cross-section is discretized into $N_f = 20$ fibers of uniform height $H/N_f$ stacked along the height. The rod swells freely and is restricted to deform in the $x$-$z$ plane. The target shape ``NU'' is illustrated in \Cref{fig:fig_result_swell_nu}(B).

To program the target deformation, we assign a design variable to every fiber of every element, resulting in $6000$ design variables $\bm{\zeta}$, while the degrees of freedom remain those of the centerline. The design variables are filtered along the length of the rod and projected to obtain the pseudodensity $\rho_{ef}$ of fiber $f$ in element $e$, where $\rho_{ef} = 1$ corresponds to gel and $\rho_{ef} = 0$ corresponds to elastomer. The shear modulus $G$ and the Flory-Huggins parameter $\chi$ of each fiber segment are then interpolated as

\begin{equation}
  G_{ef} = \rho_{ef} \, G_{gel} + (1 - \rho_{ef}) \, G_{elast}, \qquad \chi_{ef} = \rho_{ef} \, \chi_{gel} + (1 - \rho_{ef}) \, \chi_{elast}.
  \label{eq:material_interp}
\end{equation}

We adopt the shape-morphing objective in \Cref{eq:objective_morphing}, where the mismatch between the realized and target curvature is minimized. In addition to the grayness constraint, the gel volume fraction is bounded from above to a nominal value of 0.4. The optimization is performed with an MMA move limit of $5 \times 10^{-4}$.

The swollen configurations of the optimized designs for the three aspect ratios are illustrated in \Cref{fig:fig_result_swell_nu}(C), overlaid on the target shape. In all three cases, we confirm strong match between the target shape and the optimized result. For brevity, the remaining analyses in \Cref{fig:fig_result_swell_nu} are only presented for $\mathcal{A} = 200$. The optimized material layout in the undeformed configuration is illustrated in \Cref{fig:fig_result_swell_nu}(D). The curvature distributions are compared in \Cref{fig:fig_result_swell_nu}(E), which shows agreement along the entire length of the rod.  As expected, the gel is placed on the convex side of each bend, while the straight segments remain elastomer. The convergence of the objective and the constraints is illustrated in \Cref{fig:fig_result_swell_nu}(F). The optimization reduces the objective by several orders of magnitude over 250 iterations, with a final grayness of 0.02 and a gel volume fraction of 0.14. Note that the gel constraint remains inactive. Similar convergence characteristics were observed for the other aspect ratios. Finally, the evolution of the material distribution at select iterations during the optimization is illustrated in \Cref{fig:fig_result_swell_nu}(G), where the initially uniform straight rod progressively morphs towards the target. Having illustrated the framework on a planar example, we now turn to the three-dimensional design of a shape-morphing stent.

\subsection{Case II: Conforming Swelling-Programmed Stents}
\label{sec:expts_stent}

As a second case study, we consider the design of patient-specific shape-morphing stents. A stent is a mesh tube that is inserted into a vessel to counteract a disease-induced constriction of the flow \cite{tambavca2010StentDefinition}. Once deployed, a stent with low bending flexibility can permanently straighten the vessel and injure its wall \cite{kapoor2024roadToIdealStent, auricchio2013StentWallStraightening}. A direct remedy is a stent whose unloaded shape upon deployment is that of the vessel. Since the shape of the vessel differs from patient to patient, such a stent must be personalized. The personalization of stents and the use of the material distribution as a design variable have been identified as open directions in stent design \cite{kapoor2024roadToIdealStent}. Motivated by both the line of prior work and recent demonstrations of hydrogel-based shape morphing stents \cite{liu4DPrintedHydrogel2024, nagakawaHydrogelStent2022}, we seek a gel-elastomer distribution such that a stent that is straight when dry morphs to the shape of the patient's vessel upon swelling in body fluid.

\begin{figure}[]
    \centering
    \includegraphics[scale=0.7,trim={0 0 0 0},clip]{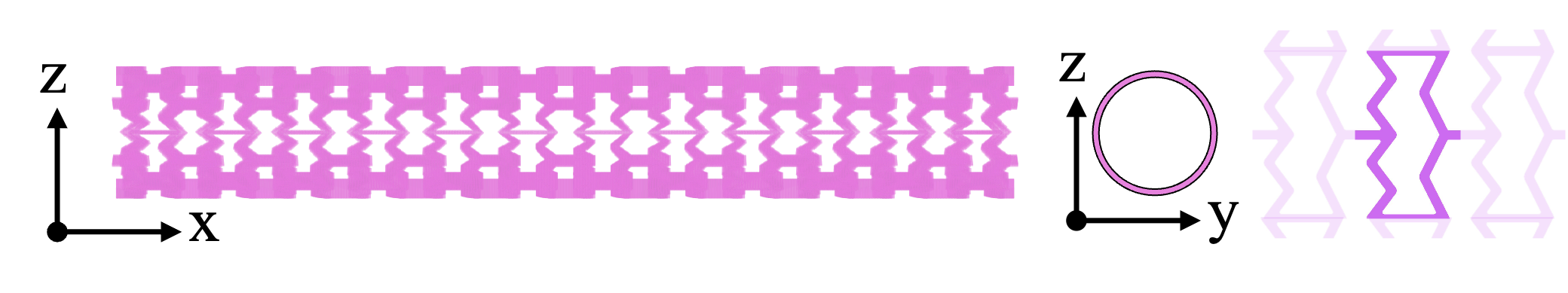}
    \caption{The stent considered in the dry state. (Left) Side view. (Middle) Cross-section. (Right) Unit cell of the Express-type lattice \cite{vcanic2023stentCrossSection}.}
    \label{fig:fig_result_stent_setup}
\end{figure}

The stent considered is illustrated in \Cref{fig:fig_result_stent_setup}. We adopt the lattice of a commercial Express-type stent \cite{vcanic2023stentCrossSection}, where the unit cell is repeated 12 times along the length and 6 times along the circumference. Note that the lattice is kept fixed, and only the material distribution is optimized. The stent has an assumed aspect ratio $\mathcal{A} \coloneqq L/R = 35$, where $L$ and $R$ denote its length and radius. Considering only the deformation of the centerline of this slender stent, we model the stent as a single Cosserat rod with an annular cross-section. The rod is discretized into $N_e = 300$ elements, and the cross-section is discretized into $N_f = 180$ fibers along the circumference, where the fiber segments that do not lie on the lattice are treated as void.

\begin{figure}[]
    \centering
    \includegraphics[scale=0.6,trim={0 0 0 0},clip]{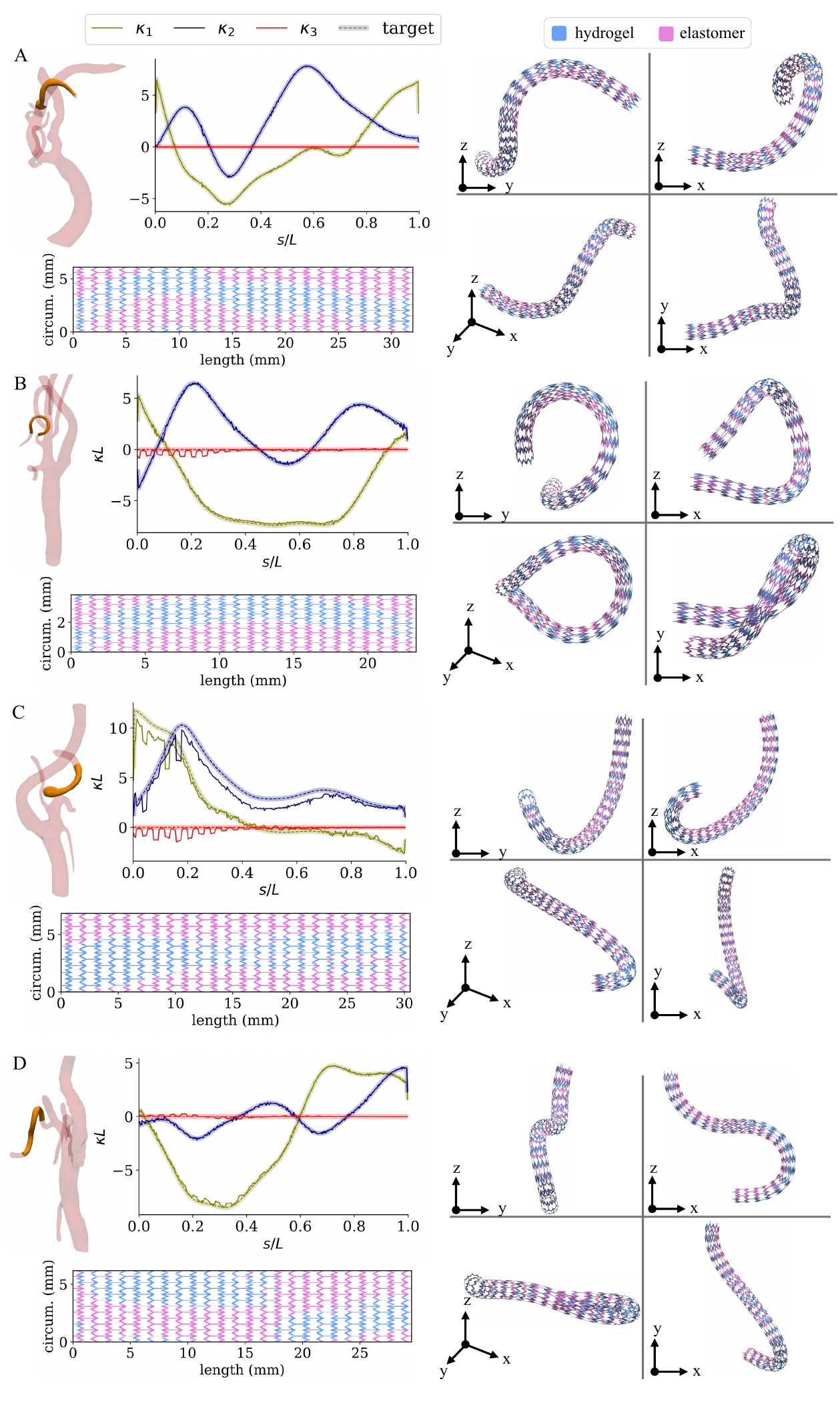}
    \caption{Patient-specific shape-morphing stents, where rows A to D correspond to four patients. For each patient: (left) the carotid vessel with the selected segment highlighted; (middle, top) the realized and target curvatures; (middle, bottom) the optimized material layout unrolled along the circumference; (right) four views of the swollen stent.}
    \label{fig:fig_result_stent_cases}
\end{figure}

The target three-dimensional shapes are derived from the carotid dataset in \cite{eulzer2024carotidDataset}, where the lumen and the centerlines of the vessels were reconstructed from computed tomography angiography scans of patients. Herein, we present the results on four selected tortuous segments, each from a different patient, as highlighted in \Cref{fig:fig_result_stent_cases}, and extract the target curvature from their centerlines. The length $L$ of each stent is set to the length of the corresponding segment, and its radius is specified by $\mathcal{A}$. Furthermore, the target is assumed to be free of twist.

As in \Cref{sec:expts_NU}, we assign a design variable to every fiber segment on the lattice, and adopt the material interpolation in \Cref{eq:material_interp} and the shape-morphing objective in \Cref{eq:objective_morphing} (\Cref{tab:case_summary}). Here, the design variables are filtered along the length with a radius of 0.5 mm, and the mismatch between the realized and target curvatures is minimized over the two bending components and the twist. Furthermore, only the grayness constraint is imposed, and the optimization is performed with an MMA move limit of $10^{-3}$.

The optimized designs for the four patients are illustrated in rows A to D of \Cref{fig:fig_result_stent_cases}. For each patient, we illustrate the vessel with the selected segment, the realized and target curvatures, the optimized material layout unrolled along the circumference, and four views of the swollen stent. Observe that the realized bending curvatures $\kappa_1$ and $\kappa_2$ closely follow the target, with some local discrepancies. In contrast to \Cref{sec:expts_NU}, the optimized layouts cannot be readily interpreted by inspection, which underscores the need for optimization. Furthermore, observe that a distinct layout is obtained for each patient on the same lattice, which highlights the patient-specific customization enabled by the framework. Convergence characteristics similar to those in \Cref{sec:expts_NU} were observed in all the cases.

We remark that the present study only targets the conformity of the stent to the shape of the vessel. Other considerations pertaining to the deployment of the stent, its post-deployment structural response such as radial strength and fatigue, and the hemodynamics \cite{kapoor2024roadToIdealStent} are not considered in the present scope, and could be future work. 
\subsection{Case III: Magnetically Steered Catheters for Reduced Wall Contact}
\label{sec:expts_catheter}

As a final case study (\Cref{tab:case_summary}), we consider the design of magnetically steered catheters for minimally invasive procedures. In endovascular procedures, a catheter is advanced through tortuous vessels to reach a target site, where it is steered primarily through its contact with the vessel wall \cite{kim2019ferromagneticSoftRobotCatheter}. However, the contact between the catheter and the vessel wall can injure the wall, and has been associated with complications such as dissection, perforation, thrombosis, and embolic stroke \cite{razban2020contactForce}. Catheters with hard-magnetic particles embedded in a soft body can instead be steered remotely through an applied magnetic field \cite{kim2019ferromagneticSoftRobotCatheter, wang2021evolutionaryDesignCatheter, pittiglio2022patientSpecificCatheterMagnetic}. Here, we seek to design the remnant magnetization of the catheter and the applied field such that the normal contact force exerted on the vessel wall during the insertion is minimized.

\begin{figure}[]
    \centering
    \includegraphics[scale=0.9,trim={0 0 0 0},clip]{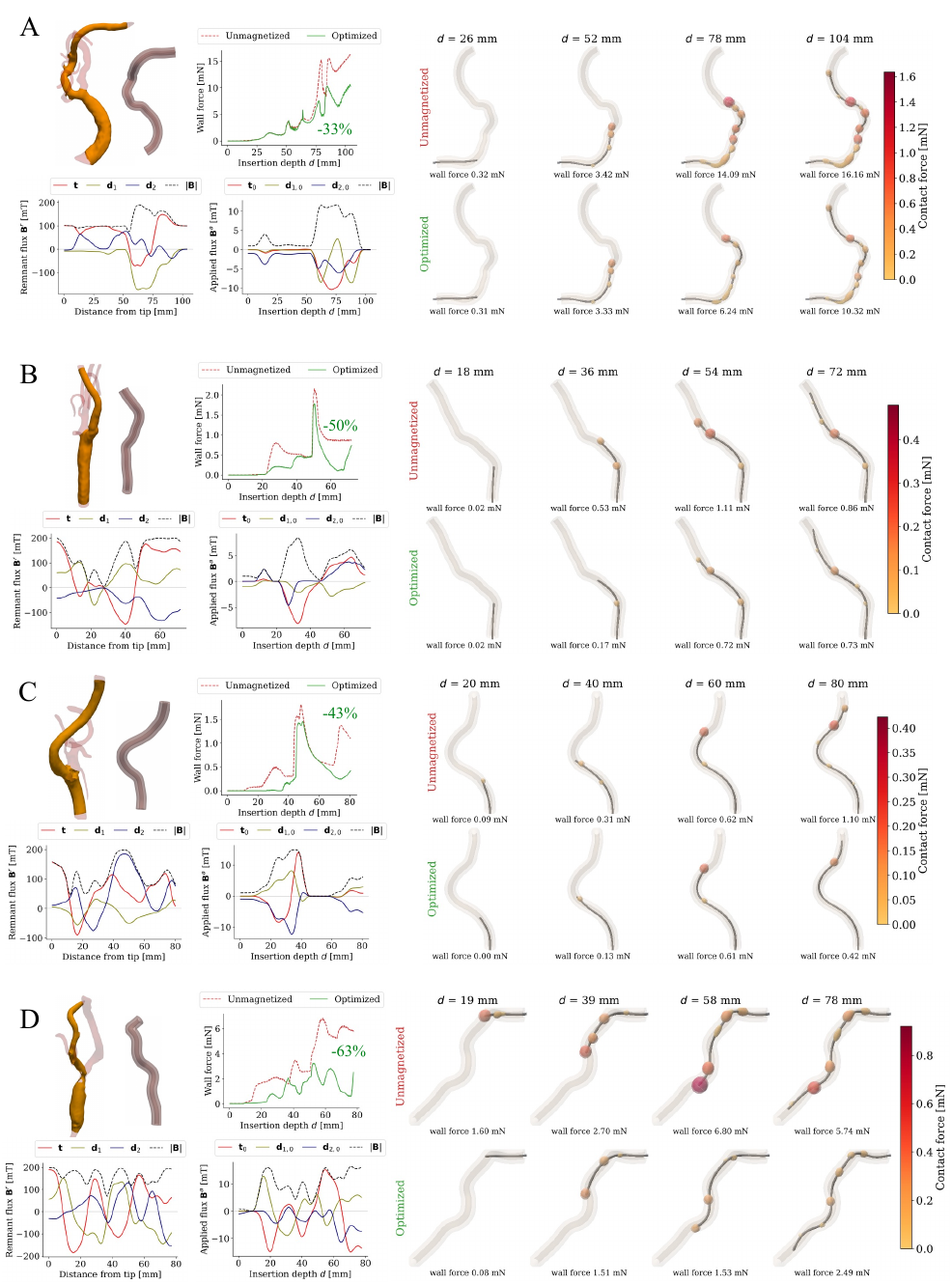}
    \caption{Magnetically steered catheters, where rows A to D correspond to four patients. For each patient: (left) the carotid vessel with the selected path highlighted, and the idealized vessel considered in the simulation; (middle, top) the wall force as a function of the insertion depth; (middle, bottom) the optimized remnant flux density along the catheter and the optimized applied flux density over the insertion, resolved along the tangent and the two directors; (right) the contact forces of the unmagnetized and the optimized catheters at 25\%, 50\%, 75\%, and 100\% of the maximum insertion depth, where the size and the color of each sphere denote the contact force.}
    \label{fig:fig_result_mag_catheter}
\end{figure}

The vessels considered are illustrated in \Cref{fig:fig_result_mag_catheter}. As in \Cref{sec:expts_stent}, the vessels are derived from the dataset in \cite{eulzer2024carotidDataset}. For each of the four patients, we extract a path along the centerline, and model the vessel as a tube with a circular lumen of radius 1.5 mm along this path. The vessel wall is assumed to be rigid and frictionless. The catheter is assumed to be a slender annular tube with an outer radius $R_o = 0.5$ mm and an inner radius $R_i = 0.35$ mm. The aspect ratio $\mathcal{A} \coloneqq L/R_o$ ranges from 175 to 270 in the cases considered, which makes it well suited to the proposed formulation targeting slender structures. The catheter is discretized into $N_e = 120$ elements, and its cross-section is discretized into $N_f = 8$ fibers along the circumference. The catheter is advanced quasi-statically into the vessel along the tangent at the entry in $K = 382$ increments. The contact with the vessel wall is modeled as described in \Cref{sec:method_contacts}, where the contact radius is set to $r_e = R_o$, the transition width to $w = 0.25 r_e$, and the penalty stiffness to $\omega_{con} = EI/\ell_{con}^4$, with $E = 3G$ and $I$ the second moment of area of the cross-section.

The design variables $\bm{\zeta}$ entail both the remnant magnetization of the catheter and the applied external magnetic field, forming a material-stimulus co-design problem~\cite{liwei2025MagneticMPM, leecodesign2026}. In particular, we assign a remnant flux density $\bm{B}^r_e$ to every element $e$, with an illustrative $\| \bm{B}^r_e \| \leq 200$ mT. Here, $\bm{B}^r_e$ is specified in the global frame in the reference configuration and is uniform over the cross-section. We also assign a spatially uniform applied flux density $\bm{B}^a_k$ to every increment $k = 1, \ldots, K$, with an illustrative $\| \bm{B}^a_k \| \leq 20$ mT. Each flux density is parameterized by its magnitude, polar angle, and azimuthal angle, resulting in $3N_e + 3K$ design variables. The remnant and applied flux densities are filtered with a radius of 5 mm along the length of the catheter and along the insertion depth, respectively. We adopt the objective in \Cref{eq:objective_contact}, where the wall force summed over all the increments is minimized. No constraints are imposed. The optimization is performed with an MMA move limit of $10^{-2}$.

The optimized designs for the four vessel segments from different patients are illustrated in rows A to D of \Cref{fig:fig_result_mag_catheter}. For each vessel segment, we illustrate the vessel with the prescribed path, the wall force as a function of the insertion depth $d$, the optimized remnant and applied flux densities, and the contact forces at four insertion depths. Here, an unmagnetized catheter, which is steered by the contact with the wall alone, serves as the reference. Observe that the optimized designs reduce the wall force over most of the insertion. In particular, the objective is reduced by 33\%, 50\%, 43\%, and 63\% in cases A to D, respectively.  The contact forces are illustrated at 25\%, 50\%, 75\%, and 100\% of the maximum insertion depth, where the size and the color of each sphere denote the contact force at that location. For instance, in case A at $d = 78$ mm, the wall force reduces from 14.09 mN to 6.24 mN. Observe also that the contact is reduced along the body of the catheter, and not only at its tip.

We remark that the objective is the wall force summed over the insertion, which does not guarantee that the wall force is reduced at every depth, a concern since perforation occurs once the local contact force exceeds a threshold \cite{perforationContactForce2011}. The minimization of the peak wall force is left for future work. Furthermore, other considerations such as the friction with the vessel wall and the deformation of the vessel wall \cite{frictionInjury2026} are left as future work.

\section{Conclusion}
\label{sec:conclusion}

This work proposes CROPS, a \underline{C}osserat-\underline{r}ods-based framework for the representation, design, and \underline{o}ptimization of \underline{p}rogrammable \underline{s}lender structures. The representation captures large translations and rotations of slender structures while discretizing only the centerline. Furthermore, the cross-section of the rod is resolved into fibers, allowing the material to vary both along the length and across the cross-section. Two forms of actuation are considered, namely the swelling of gel-elastomer composites and the magnetic actuation of hard-magnetic composites. The material layout, together with the applied actuation, programs the response of the structure. Additionally, the contact of the rod with itself and its surroundings is accounted for. Finally, the framework is implemented in an end-to-end differentiable manner, with the sensitivities computed via automatic differentiation, enabling the gradient-based optimization of the material layout and the actuation. We demonstrate the framework for patient-specific programmable devices. First, stents are programmed to morph to the shape of carotid vessels upon swelling. Second, the remnant magnetization and the applied field of magnetically steered catheters are co-designed to reduce the force exerted on the vessel wall during insertion.

Several directions remain open to extend this further. The fidelity of the contact model can be improved by incorporating friction, deformable contact bodies, and the interaction with a surrounding fluid \cite{gazzola2018CosseratRods}. Furthermore, the design formulation can be extended. For instance, the peak wall force can be minimized in place of the summed wall force, and stent requirements such as radial strength and fatigue \cite{kapoor2024roadToIdealStent} can be imposed alongside the target shape. Similarly, applied fields with spatial gradients and closed-loop control of the applied field \cite{francescon2025closedLoopControlCatheter} would offer finer control over the catheter. Additionally, extending the framework to assemblies of multiple rods and to dynamics \cite{parthasarathy2026elastica++} would broaden its scope to untethered soft robots \cite{truby2024HSASoftRobot}. Finally, the fabrication and experimental testing of the optimized designs \cite{liu4DPrintedHydrogel2024, kim2019ferromagneticSoftRobotCatheter} is essential to validate the framework. Ultimately, CROPS provides a differentiable platform for the design of programmable slender structures, with direct applicability to minimally invasive devices, soft robotics, and deployable structures.


\section*{Declaration of Competing Interest}

The authors declare that they have no known competing financial interests or personal relationships that could have appeared to influence the work reported in this paper.

\section*{Compliance with ethical standards}
The authors declare that they have no conflict of interest.

\section*{Acknowledgments}
A.C. and W.C. acknowledge the support of DARPA METALS under agreement No. HR0011-24-2-0302. W.C., D.D.L., S.L. and H.K. also acknowledge support from the NSF Boosting Research Ideas for Transformative and Equitable Advances in Engineering (BRITE) Fellow Program (CMMI-2227641). W.C. and  D.D.L. are also grateful for support from NASA's Minority University Research and Education Project Institutional Research Opportunity (MIRO) through the Center for In-Space Manufacturing (CISM-R2): Recycling and Regolith Processing (Award 80NSSC24M0176). W.C. acknowledges support from the NSF Materials Research Science and Engineering Centers (MRSEC) program (DMR-2308691).

\section*{Declaration of generative AI and AI-assisted technologies}

During the preparation of this manuscript, the authors used Google’s Gemini and Anthropic’s Claude to improve the language and readability. After using these tools, the authors reviewed and edited the content as needed and take full responsibility for the content of the publication.

\section*{Replication of Results}
The Python source code is available at \href{https://github.com/ideal-nu/crops}{github.com/ideal-nu/crops}.

\bibliographystyle{unsrt}  

\bibliography{7_references}

\end{document}